# Human Agency and Creativity in AI-Assisted Learning Environments

Yun Dai

The Chinese University of Hong Kong

**Abstract**

This chapter explores human creativity in AI-assisted learning environments through the lens of student agency. We begin by examining four theoretical perspectives on agency—instrumental, effortful, dynamically emergent, and authorial—and analyze how each frames the relationship between agency and creativity. Under each theoretical perspective, we discuss how the integration of generative AI (GenAI) tools reshapes these dynamics by altering students' roles in cognitive, social, and creative processes. In the second part, we introduce a theoretical framework for agentic engagement in AI-assisted learning, contextualizing agency within specific cognitive, relational, and ethical dynamics introduced by GenAI tools. This framework is linked to the concept of Mini-c creativity, emphasizing personal relevance and self-directed learning. Together, these perspectives support a shift from viewing creativity as product-oriented to understanding it as a process of agentive participation and meaning-making. We conclude with two future research directions focused on the creative process and performance in AI-assisted learning.

**Introduction**

The rise of generative artificial intelligence (GenAI) in education has sparked sustained attention to student agency and creativity. As students have increasingly adopted GenAI tools for academic tasks, questions arise about how these technologies influence students' capacity to make choices, take ownership of their learning, and express original ideas. While GenAI tools offer immediate, human-like outputs in various forms (e.g., text, images, audio, and code), they can also narrow the space for student-driven inquiry and limit authentic creative expression. Some students may rely on AI-generated outputs, which can reduce their engagement in critical thinking and self-directed exploration. Others may feel uncertain about when to trust their own ideas versus those of GenAI outputs. Such AI-assisted learning experiences, as well as the ways in which students interact with GenAI, have significant implications for the cultivation of agency and creativity. Against this background, this chapter addresses these complex dynamics between student agency and creativity in AI-assisted learning environments.

Nevertheless, student agency has long been viewed as a tricky concept due to the difficulty in defining, operationalizing, and measuring it, as well as a lack of consensus among researchers (Archer, 2000; Hitlin & Elder Jr, 2007). Meanwhile, the idea of agency has been increasingly shaped by *neoliberal ideologies*, which tend to portray individuals as fully autonomous, highly adaptable, and responsible for their own success—as if they are entrepreneurs managing their own lives (Ganti, 2014). This neoliberal framing can be problematic in education, as it assumes full responsibility on students and teachers without considering social, structural, or institutional constraints (Gershon, 2011). To address this problem, researchers and educators need to go beyond simplistic or idealized views of agency. Instead, we should critically examine the different ways agency is expressed and work toward developing a normative understanding of what agency means and how student agency is related to creativity in educational contexts.

This chapter explores student agency as a central lens for understanding human creativity in AI-assisted learning environments. Given the conceptual ambiguity surrounding agency, it is essential to begin with a clear framework that clarifies how student agency is theorized within educational research. To this end, we adopt the typology proposed by Matusov, von Duyke, et al. (2016), which outlines four key perspectives on student agency: instrumental, effortful, dynamically emergent, and authorial agency. Using this framework, we examine how the integration of GenAI tools is reshaping student agency and creative expression in educational research and practices. Building on this foundation, we propose a person-centered approach that integrates students' agentic engagement with mini-c creativity and call for research that examines how students initiate, negotiate, and sustain creative actions within the affordances and constraints of AI-mediated educational settings.

**Conceptualization of Student Agency in Educational Studies**

Matusov, von Duyke, et al. (2016) synthesize four theoretical perspectives on student agency in educational discourse: instrumental, effortful, dynamically emergent, and authorial agency. These perspectives not only frame how agency is conceptualized but also carry significant implications for understanding and practicing creativity in educational settings. In this section, we examine each form of agency in turn, analyzing how it relates to student creativity and how the recent integration of GenAI tools impacts these dynamics. Table 1 provides an overview of the four theoretical perspectives. Through this structured framework, we synthesize an intersectional body of theoretical arguments to provide an overview of how student agency and creativity are conceptualized, examined, and debated within the context of AI-assisted learning environments.

**Table 1**

*An Overview: Conceptualizations of Student Agency and Its Implications for Creativity and AI-assisted Learning Engagement*

| Category | Definition of Agency | Interplay with Creativity | Implications for AI-assisted Creative Learning |
|---|---|---|---|
| **Instrumental** | Individual's capacity to rationally achieve predefined goals through autonomous, goal-directed actions. | • Provides raw materials for creative tasks.<br>• May kill creativity due to rigid curricula and standardized assessments.<br>• Limits exploratory and unpredictable aspects of creativity. | • Supports efficient task completion and skill acquisition.<br>• Enables rapid iteration and fluency in basic idea generation.<br>• Risks promoting superficial creativity focused on task performance rather than open-ended exploration. |
| **Effortful** | Individual's sustained motivation, persistence, and personal commitment to pursue and complete predefined tasks. | • Fosters creativity when driven by intrinsic motivation and personal relevance.<br>• Supports sustained creative effort and deep engagement.<br>• May limit creativity if driven by external pressure or rigid expectations. | • Assists with cognitive load reduction, allowing students to focus on higher-order thinking.<br>• Can sustain engagement through iterative feedback and support.<br>• Risks reducing meaningful effort if used as a shortcut for ideation or task execution. |
| **Dynamically Emergent** | Agency emerges from interactions among people, environments, and materials as a relational and distributed processes. | • Creativity as collaborative and co-constructed.<br>• Creativity emerges through interaction with peers, tools, and environments.<br>• Shifts focus from individual outcomes to collective innovation. | • Positions GenAI as a distributed cognitive partner in co-creative processes.<br>• Encourages collective idea generation and knowledge building and promote new models of networked creativity.<br>• Risks dilute individual responsibility and depersonalize agency. |
| **Authorial** | Individual's capacity to transcend given cultural and social norms through critical, dialogic engagement and responsible authorship of new meanings and practices. | • Frames creativity as personal authorship and cultural transformation.<br>• Encourages critical interrogation and ethical reflection.<br>• Values originality emerging from students' unique voices and social commitments. | • Treats GenAI as a source of cultural material for critical and creative reinterpretation.<br>• Promotes dialogic engagement with GenAI outputs rather than passive consumption.<br>• Resists standardization and predictability, making creative contributions difficult to evaluate within institutional frameworks. |

*Instrumental Agency*

Instrumental agency emphasizes the capacity of individuals to achieve predefined goals through rational, autonomous action (Edwards & Mackenzie, 2005). Rooted in Enlightenment thought and rationalism, it views agency as a tool to produce rational outcomes, often in individualistic and independent manners (Campbell, 2009). Under an instrumental view, goals are defined in advance of action, and the agent's job is to carry out actions to achieve those goals, regardless of the person's own purposes and contextual meanings. Instrumental agency is pervasive in both formal education and organizational settings. In education, it fits technocratic, outcomes-based models, where success is measured by predefined objectives, standardized skills, and quantifiable outcomes (Matusov, 2011). Within this framework, curriculum standards, testing, and the "skill-and-drill" methods dominate—students are expected to acquire a fixed "toolkit" of knowledge and skills and apply it to achieve externally educational goals.

While the instrumental view of student agency offers some benefits, it is more often criticized for constraining rather than fostering creativity (Wahl, 2022). This approach emphasizes the acquisition of predefined skills and competencies, which can provide a necessary foundation for creative work—such as mastering language conventions for writing or technical skills for artistic and scientific projects. However, when schooling becomes overly instrument-driven—fixated on standard curricula, testing, and predefined learning outcomes—it tends to suppress imaginative exploration, experimentation, and risk-taking (Beghetto, 2005). In practice, the pressure to meet standards and cover content can lead teachers to adopt rigid pedagogies at the expense of creativity. As for students, when external goals (test scores, teacher approval, grades) dominate, they may learn that their own curiosities and ideas are secondary, eventually dampening their creative thinking (Beghetto & Kaufman, 2014). The key tension is that creative learning is often *non-linear* and *exploratory*, whereas instrumental education seeks efficiency and clarity of progression in individuals.

The integration of GenAI adds a new layer of complexity to the relationship between instrumental agency and student creativity. On the one hand, GenAI tools—trained on vast datasets—offer students immediate access to diverse linguistic, conceptual, and disciplinary resources. These tools can support early stages of creative work by providing raw material for brainstorming, exploring alternatives, and refining drafts (Wang et al., 2023). From an instrumental perspective, GenAI may appear to enhance student efficiency in achieving externally defined tasks. However, this very efficiency challenges the foundational logic of instrumental agency. When students can bypass traditional processes of memorization, procedural practice, or stepwise skill acquisition, the rationale behind "skill-and-drill" pedagogy is inevitably weakened (Yun Dai et al., 2023). In this context, the instrumental view of agency becomes less applicable, as the goals themselves begin to lose relevance. Emerging research also raises concerns that AI assistance may reduce students' active engagement and ownership of their work, thereby weakening their sense of agency (Darvishi et al., 2024).

### *Effortful Agency*

Effortful agency refers to the capacity of individuals to initiate, sustain, and take responsibility for actions (Campbell, 2009). Unlike instrumental agency, which emphasizes rational competence, effortful agency is characterized by personal motivation, the will to overcome obstacles, and a committed investment of effort. It highlights the internal work required to persist with actions that may not originate from the individual but are taken up as meaningful through sustained engagement. Rather than treating agency as a fixed trait or automatic response, this perspective sees it as an ongoing achievement shaped by the interplay between social influence and personal affirmation (Bandura, 2001). The individual does not merely comply with expectations but actively appropriates them, imbuing their actions with intention and accountability. A representative form of effortful agency is motivation, particularly when understood as a dynamic process of internalizing and sustaining purposeful action (Franzese, 2013). In fact, many studies have used "agency" and "motivation" interchangeably, reflecting the central role of volition and commitment in how individuals undertake and carry through with socially structured tasks.

The relationship between creativity and motivation has been well studied and documented. From this motivation perspective, creativity is not simply the outcome of individual ability, but a function of students' capacity to initiate and direct their own cognitive and creative processes (Prabhu et al., 2008). Empirical research over the last 20 years generally supports the idea that intrinsically motivated students tend to exhibit higher creativity, while certain extrinsic motivators, such as expected evaluation or reward, can "kill" creativity by shifting students' focus away from exploration toward narrow goal attainment (Collins & Amabile, 1999; Hennessey, 2012). Although the empirical evidence is compelling, keep in mind that not all forms of motivation neatly conform to this simple equation—context and individual differences play a significant role. Research has identified several key factors that can influence the strength and nature of the relationship between motivation and creativity in students. These include aspects of the learning environment and teaching practices (Yuan et al., 2019), cultural context (Eisenberg, 1999), and other individual or situational variables (Sung & Choi, 2009) that moderate or mediate the relationship between motivation and creativity.

The impact of GenAI on student creativity, viewed through the lens of effortful agency, emerges as a focal point in recent research (Holzner et al., 2025). GenAI tools, such as ChatGPT and other content generators, have been rapidly adopted in educational and professional settings as interactive partners for brainstorming, ideation, and other cognitive explorations (Wang et al., 2023). Rather than replacing student effort, GenAI can support learners by reducing knowledge gaps, providing immediate feedback, and enabling them to work more independently. This assistance allows students to experiment with ideas, refine their thinking, and engage in creative tasks without overreliance on external guidance (Y. Dai et al., 2023). It can also reduce student frustration and negative emotions often associated with the iterative and uncertain nature of creative work. In this regard, by helping students stay within a productive zone of challenge, GenAI fosters persistence and a growing sense of competence and control. These experiences could create a positive feedback loop that strengthens effortful agency and enhances their creative performance (Zhou & Peng, 2025).

While GenAI offers meaningful support, it also presents significant challenges. A central concern is the potential overreliance on AI tools, which can diminish students' active engagement in the creative process (Zhai et al., 2024). Because GenAI systems are trained on extensive datasets, they tend to generate responses that reflect common patterns rather than truly novel or divergent ideas (Doshi & Hauser, 2024). When students accept these outputs at face value—rather than critically questioning, revising, or building upon them—they risk limiting their creative exploration (Zebua, 2024). This reliance can also lead to what some scholars describe as *metacognitive laziness*, where learners defer critical thinking and self-reflection in favor of immediate, passable answers provided by AI (Fan et al., 2025). From the perspective of Self-Determination Theory, the passive use of AI may erode students' sense of autonomy and competence—two foundational needs for intrinsic motivation (Deci & Ryan, 2012). Such misuse can turn learning into an AI-based routine, diminishing students' creative engagement as well as their ownership of the exploratory process.

The dual role of GenAI in student creativity can be understood as a tension between augmentation and dependence. Examining the factors that mediate the augmentation or dependence results has become a central focus in learning sciences. Consistent with findings from non-AI contexts, some early studies show that intrinsically motivated students are more likely to engage critically and reflectively with GenAI tools to enhance their creative performance, while extrinsically motivated students tend to use them to shortcut tasks and minimize cognitive effort (Singh & Aziz, 2025; Ye et al., 2025). Other factors, such as AI literacy, trust or reliance on AI, critical thinking, and self-regulated learning, have also been identified as important. For example, Lee et al. (2025) have found that higher confidence in GenAI correlates with reduced critical thinking, whereas higher self-confidence is associated with more critical and reflective engagement. This suggests that both task- and person-specific factors influence creative engagement, but through different pathways. However, contrasting evidence from Lijie et al. (2024) found no mediating effect of motivation between critical thinking and creativity, despite a strong direct effect of critical thinking on creative performance. These mixed findings highlight the complexity of this emerging field and the need for further empirical research to disentangle the interactions among motivational, cognitive, and contextual factors in AI-assisted creativity.

### *Dynamically Emergent Agency*

A dynamically emergent approach views agency as arising from complex, often unpredictable interactions among individuals, environments, and materials (Emirbayer & Mische, 1998; Goller & Goller, 2017). It emphasizes holism, materiality, and distributed action, shifting the focus from individual intention to emergent systems in which agency may be distributed to collectives, technologies, or environments (Larsen–Freeman, 2019; Wertsch & Rupert, 1993). This perspective suggests that creativity can emerge spontaneously through interaction rather than deliberate design, offering valuable insights into how material conditions and contextual affordance shape action, particularly in ecological and self-organizing educational models (Leonardi, 2012; Wertsch et al., 1993). However, critics argue that it risks detaching agency from personhood, responsibility, and dialogue, reducing it to a depersonalized force lacking ethical accountability (Matusov, von Duyke, et al., 2016). While debates continue over whether agency must be anthropocentric, this approach foregrounds structural and contextual influences on creative action.

To address the issues, Anne Edward and others (Edwards & D'arcy, 2004) propose the notion of relational agency, which refers to the capacity to work with others and negotiate goals dynamically. Behind this notion is the idea that creativity often emerges in collaborative contexts, and thus, agency should be thought of not only at the individual level but also at the group or network level (Valquaresma, 2024). For example, in a design-thinking course, a team of students might brainstorm together; no single student controls the outcome, but collectively they arrive at an innovative solution. Traditional instrumental views sometimes neglect this by isolating individual performance. New pedagogies like knowledge-building classrooms (Scardamalia & Bereiter, 2006, 2021) treat the class as a knowledge-creating community where students share ideas freely and build on each other's contributions via technologies. It is believed that when students experience this kind of joint agency, the ideas generated can surpass what any one student would have developed alone, indicating a social dimension to creativity.

The rapid advancement of computational and intelligent technology has extended the distributed, relational agency into the realm of *human-technology co-agency*. In human factors and Human-Computer Interaction (HCI), long-standing research on human-automation teaming highlights the interdependence between human operators and machines (O'neill et al., 2022; Rammert, 2008). Woods and Hollnagel (2006) describe this interdependence as "joint cognitive systems," in which humans and machines collaborate to perform tasks. This perspective reflects a shift away from anthropocentric views of agency, recognizing that agency can also emerge from material and technological entities (Knappett & Malafouris, 2008). Human–technology coagency refers to situations where humans and technological entities (such as AI systems, machines, or digital tools) jointly share and enact agency in pursuing goals. Rather than serving merely as passive instruments, technologies actively shape outcomes through their affordance and actions, forming a co-agentic partnership with human actors.

The coagency perspective has also influenced the research on creativity, especially in creative arts field: creativity was long viewed as a uniquely human domain, but increasingly artists and designers are experimenting with GenAI tools (like generative art algorithms, music composition AIs, or large language models for writing) as partners in the creative process

(Nordström et al., 2023). Research on human–AI *cocreativity* finds that agency is truly distributed in these collaborations: the AI contributes ideas or raw generative outputs, and the human guides, curates, or modifies them, often in an iterative loop (Bown, 2015; Loivaranta et al., 2025). The creative agency lies in the dialogue between human and machine—the AI's "suggestions" can inspire the human to explore approaches they wouldn't have conceived alone (Huang, 2023). Lim et al. (2023) further conceptualize GenAI as a "relational artefact" that can spur transformative creativity in learners, when perceived as an empathetic, collaborative partners. However, the coagency also raises concerns about diminished human authorship, as strong AI influence may obscure the human voice and prompt users to undervalue AI-generated contributions, despite their active role in guiding the process (Loivaranta et al., 2025).

Educators and researchers are now exploring human–AI coagency as a partnership, where GenAI tools function as an active collaborator that can support and augment student learning and creativity (Ibrahim, 2024). Research suggests that human-AI collaboration can shape both the creative process and product (Cui et al., 2024; Kim & Maher, 2023). GenAI, in particular, is effective at stimulating ideation, helping students move beyond creative blocks and consider novel possibilities. In an experimental study by Hwang and Lee (2025), students who engaged in iterative prompt refinement with AI demonstrated significantly stronger creative problem-solving skills than those who did not. They described the AI as a "valuable partner" for idea generation, enhancing both the speed and diversity of thinking.

Despite early promising evidence, coagency in education calls for cautious and ethical challenges. Jia et al. (2024) found that its positive impact on creativity is more pronounced among higher-skilled individuals in professional settings. Zhou and Peng (2025) reported that AI use in teaching correlates with increased student creativity, primarily mediated by engagement. That is, when students are more engaged through AI-driven personalization, they invest more in creative thinking and output. Meanwhile, coagency also raises concerns for educators. A qualitative study by Mouta et al. (2025) found that K–12 teachers were concerned about the intersubjective and collective dimensions of agency in classrooms, particularly the risk that AI-driven instructional decisions could undermine shared authority and diminish the teacher's role. This tension highlights a key challenge: while AI can empower learners through autonomy and feedback, it also requires renegotiating the agency of teachers and students. The goal is to strike a balance where AI augments rather than displaces human intentionality, preserving genuine co-creation in learning.

*Authorial Agency*

Authorial agency refers to an individual's capacity to create meaning, culture, and action that goes beyond established social norms and practices (Matusov, von Duyke, et al., 2016). It is grounded in the idea that agency is not simply about following established goals or complying with given structures, but about engaging with the "given" as material for personal and socially recognized transformation (Matusov, Smith, et al., 2016). Grounded in sociocultural theories, authorial agency is seen as dialogic, evaluative, and ethically anchored. It emerges through mutual recognition and carries responsibility for the consequences of one's actions. Importantly, it treats action as *praxis*—the construction and transformation of values and goals through action itself—rather than *poiesis*, which implies predefined outcomes (Boaler & Greeno, 2000; Cimasko & Shin, 2017). Greeno (2006) emphasizes that developing agency requires opportunities to participate in interactions where individuals are positioned as accountable authors of their actions. This perspective challenges reproduction-oriented models of education and instead affirms personal authorship and unpredictability as central to meaningful learning and action.

Authorial agency is deeply connected to student creativity, particularly in the forms of critical thinking, innovation, and cultural production (Baxter Magolda, 2007). Creativity, in this view, is not about following procedures to generate novel outputs, but about making original contributions that reflect the student's personal stance, voice, and responsibility (Matusov, von Duyke, et al., 2016). By transcending the given, students exercising authorial agency do not simply accept problems as defined but reframe them, challenge their assumptions, and propose new possibilities—core aspects of critical and creative thought. However, authorial agency also raises important ethical and educational questions about the direction and consequences of creative authorship. Without critical engagement, creativity can be exercised in ways that reproduce harmful ideologies or reinforce oppressive structures. The value of what is authored cannot be assumed; it must be examined in relation to its social, cultural, and ethical implications (Tappan, 1999). Thus, authorial agency requires not only the freedom to create but also a commitment to reflect on the meaning, impact, and responsibility of one's creative actions.

In the context of AI-assisted learning, the authorial agency perspective challenges its use as a mere tool for efficiency or performance optimization. Instead, it emphasizes the potential for students to use GenAI to *author* their own expressions, arguments, designs, and inquiries. A growing body of empirical studies examines how GenAI facilitates creative expressions in both educational and professional contexts (Wagler, 2024). These tools enable students to produce multimodal work—text, images, and music—often beyond their technical skills, helping bridge expertise gaps in creative production (Haase et al., 2023; Hitsuwari et al., 2023). For example, engineering students have used GenAI in their product design courses to create flyers and promotional materials that would otherwise require advanced design expertise (Dai, 2025). Other studies focus on authorship in AI-assisted writing (e.g., Formosa et al., 2025). Singh et al. (2025) found that a specifically designed intelligent tutoring system significantly improved both the authorial voice and authorship beliefs of English as a Second Language (ESL) students in Fiji. Together, these findings suggest that when designed and integrated thoughtfully, GenAI can serve as a scaffold for developing authorial agency, especially among learners with limited access to traditional expressive tools.

The authorial agency perspective also allows us to critique the status quo of studies examining the impact of GenAI assistance on creativity. Most studies still focus on measurable outcomes such as novelty, fluency, or efficiency, overlooking the deeper processes of meaning-making, authorship, and transformation. Few empirical studies—especially in-depth, qualitative accounts—have examined how students use GenAI to construct personally meaningful expressions or inquiries. From an authorial agency lens, creativity is not simply the generation of new ideas but the extent to which learners take ownership of AI-assisted outputs, reflect on their significance, and position themselves dialogically in relation to the tool and the broader sociocultural context (Matusov, von Duyke, et al., 2016). This perspective highlights a critical gap in current research: the need to investigate how GenAI can support the development of creative agency, not just creative products. Future studies should explore how students negotiate authorship and identity through sustained interaction with GenAI in educational settings.

The discourse surrounding authorship, creativity, and originality also constitutes major ethical challenges in AI-assisted learning or writing environments. As GenAI tools become more capable of generating fluent, persuasive content, concerns have grown over authorship integrity, student originality, and academic misconduct (Al-Kfairy et al., 2024; Sharma & Panja, 2025).

Scholars argue that relying on AI to produce work without disclosure undermines intellectual honesty, as the labor and voice attributed to the student may, in fact, belong to the machine (Bozkurt, 2024). While AI can support idea generation and reduce barriers to expression, over-reliance risks diluting students' authorial agency and weakening critical thinking (Zhai et al., 2024). Legal and institutional frameworks have yet to fully address questions of ownership, attribution, and accountability. At the same time, educators and researchers highlight the potential for ethical, pedagogically grounded AI use—if students are positioned as active curators and interpreters of AI outputs (Higgs & Stornaiuolo, 2024). Moving forward, ethical integration requires transparency, revised assessment designs, and student engagement with the norms and responsibilities of authorship.

In sum, these four perspectives are influenced by two divergent traditions: *positivism* and *social constructionism*. The positivist tradition emphasizes objectivity, rationality, and measurable outcomes, aligning closely with instrumental and effortful notions of agency. Under positivism, student agency is viewed primarily as an individual capacity—either as competence in achieving externally defined goals or as persistent effort toward those goals (Franzese, 2013). GenAI tools within this approach are valued for their efficiency, skill-building capabilities, and precise evaluation of student performance. In contrast, the social constructionist tradition emphasizes meaning-making through social interaction and cultural context, resonating with dynamically emergent and authorial views of agency (Lasky, 2005; Wertsch et al., 1993). From this perspective, agency is understood as a relational and dialogic process where creativity emerges through collaborative interactions, critical reflection, and ethical dialogue. GenAI tools are viewed not merely as instruments for task completion, but as co-creative partners that mediate students' authorship and social interaction. Thus, positivism foregrounds individual agency and measurable creativity, while social constructionism highlights collective engagement, contextual responsiveness, and culturally situated creativity.

**Linking Agentic Engagement with Mini-C Creativity in AI-Assisted Learning**

Building on the theoretical foundations outlined above, this section proposes future research agendas that link student agency and creativity in AI-assisted learning environments. We argue for a *person-centered* approach that emphasizes agentic engagement—students' proactive control over their learning and GenAI interactions—as a key mechanism for understanding creativity. Rather than treating agency or creativity as fixed traits or decontextualized skills, we conceptualize them as situated, emergent processes that unfold through students' active participation in meaningful learning contexts. In particular, we focus on mini-c creativity—the personal, often subtle, creative insights that emerge in the process of learning—as an anchor for grounding student agency. From this view, creativity is not an extraordinary act but a process of making sense, expressing ideas, and constructing meaning in ways that are novel, valuable, and relevant to the individual. By foregrounding agentic engagement and mini-c creativity, we call for research that examines how students initiate, negotiate, and sustain creative actions within the affordances and constraints of AI-mediated educational settings.

### *Agentic Engagement in AI-assisted Learning Environments*

While Matusov, von Duyke, et al. (2016) offer a general framework for synthesizing educational discourses on student agency, the unique affordance of GenAI tools and the nature of human-AI interactions call for a more specific theoretical framework for approaching learner agency. Specifically, there is a need to conceptualize and operationalize agency in ways that capture how students initiate, regulate, and reflect on their engagements with GenAI tools (Yang et al., 2024). We are particularly interested in the perspective of effortful agency, which emphasizes agency as a self-initiative, intentional, and sustained process. However, existing research on effortful agency often operationalizes it through proxy constructs such as self-efficacy, self-determination, intrinsic motivation, or satisfaction of basic psychological needs (Mameli & Passini, 2019). While these constructs are theoretically grounded and supported by validated measures, they are mostly developed in traditional classroom environments, with a focus on intrapersonal and interpersonal interactions (i.e., interactions with selves, peers, and teachers). They tend to overlook the *digital materiality* of GenAI tools—that is, the ways in which technological systems shape, mediate, and co-construct human actions and learning processes *in situ* (Pink et al., 2017).

Given the evolving dynamics of human-AI interactions, there is a need for a refined framework that captures the specific forms of agentic behavior students enact during creative collaboration with GenAI. In AI-assisted environments, agency involves more than belief or intention—it requires the active regulation of engagement with a non-human collaborator that both enables and constrains learning (Yang et al., 2024). Student agency thus becomes a dynamic, interactional process unfolding at the micro level, which is visible in how they frame prompts, revise inputs, and respond to AI outputs. This conceptual complexity aligns with Reeve's theory of agentic engagement (Reeve, 2013; Reeve & Shin, 2020), which situates agency within the context of student engagement. He extends traditional models of engagement by adding a fourth dimension to the cognitive, behavioral, and affective components (Fredricks et al., 2004; Pineda-Báez et al., 2019). Reeve (2012) conceptualizes agentic engagement as students' proactive efforts to shape and control their learning environments. Its emphasis on initiative, co-construction, and transactional involvement resonates with the demands of AI-assisted learning, where students must actively guide and regulate their interactions with GenAI tools (Yun Dai et al., 2023; Giray, 2023). This conceptualization offers a highly relevant lens for examining student agency in dynamic, technology-mediated learning environments.

Guided by the agentic engagement theory, Dai and Lai (2025) propose a theoretical framework of student agency in AI-assisted learning through a grounded theory study. Specifically, this framework theorizes student agency in terms of four interrelated and iterative components:

- **Initiate and (re)direct**: Students purposefully instruct AI tools and adapt their instructions iteratively, ensuring that AI outputs align with their learning goals, needs, and interests. They exercise deliberate control over the tool and demonstrate proactive leadership by shaping the direction of human-AI interaction.
- **Mindful adoption**: Students critically evaluate the quality, relevance, and credibility of AI-generated content and selectively incorporate it into their learning. This reflects context-sensitive judgement and responsible decision-making in using AI to support academic goals.
- **External help-seeking**: Students proactively seek additional sources of help to validate, enrich, or extend AI-generated content. This includes consulting online resources or engaging in collaborative inquiries with instructors, peers, or other knowledgeable individuals to deepen understanding and ensure accuracy.
- **Reflective learning**: Students engage in ongoing reflection and self-scrutiny during and after their interactions with AI tools. Through this process, they assess the effectiveness of their AI use, identify areas for improvement, and intentionally refine their strategies to enhance their AI-assisted learning in a self-directed and purposeful way.

This agentic engagement framework offers a heuristic foundation for understanding student creativity in AI-assisted learning environments. Creativity in education is increasingly seen not only as the production of novel ideas, but as a process shaped by interaction, context, and intentionality. In AI-assisted settings, creativity emerges through students' active negotiation with the tool's affordances and limitations. The framework foregrounds agency as a dynamic, iterative process through which students initiate, evaluate, adapt, and reflect—practices that are essential for creative thought and action. Rather than treating creativity as an individual trait or outcome, this perspective situates it within the evolving interplay between the learner and AI, where meaning is co-constructed through purposeful engagement. By conceptualizing agency as both situated and strategic, the framework helps explain how students exercise creative control over their learning trajectories, navigating uncertainty, generating alternatives, and shaping AI use in ways that reflect their goals, values, and intellectual curiosity.

*Linking Agentic Engagement with Creative Process and Performance*

The theoretical framework of agentic engagement in AI-assisted learning resonates with the trend of Mini-c creativity study in education. Mini-c creativity refers to personally meaningful and novel interpretations learners generate as they interact with new experiences and concepts (Kaufman & Beghetto, 2009). Unlike "Big-C," which represents groundbreaking, historically recognized creativity, and "Little-c," which encompasses everyday creativity appreciated by others, Mini-c creativity is internal, subjective, and personally significant, not necessarily acknowledged externally (Beghetto & Kaufman, 2007, 2015). The concept emphasizes learners' internal sense-making and the creative insights they gain during the learning process. Similarly, agentic engagement highlights students' proactive efforts to shape and personalize their educational contexts, reflecting active, internally driven participation in learning (Reeve, 2013). Both frameworks underscore learners' self-directed contributions—whether interpreting new information through personally meaningful insights (mini-c) or initiating and guiding interactions with AI tools (agentic engagement). Thus, integrating agentic engagement with Mini-c research could enrich our understanding of how students creatively leverage their agency to co-construct personalized, meaningful experiences in AI-assisted environments.

Informed by the integrated theoretical lenses of agentic engagement and Mini-c creativity, we propose a visual model, as represented in Figure 1, to illustrate the co-construction of creativity between students and GenAI tools within AI-assisted learning environments. At the center of the model is the dynamic interaction between student and AI, represented as a reciprocal process shaped by factors on both sides. On the student side, agentic engagement is influenced by a constellation of individual-level factors—including cognition, emotion and motivation, personal disposition, and AI-related beliefs and competencies—that shape how students initiate, regulate, and reflect on their interactions with AI. On the AI side, four key dimensions—generative capability, responsiveness and adaptivity, bias and constraint, and interface and design features—define the system's affordances and limitations. These human and technological elements are situated within broader environmental contexts, including classroom, institutional, and sociocultural influences, all of which converge to mediate the creative process. Together, the model conceptualizes creativity not as a product of isolated human or machine activity, but as a socially and materially situated process of agentic co-construction.

**Figure 1**

*A conceptual model of student–AI co-construction of mini-c creativity in AI-assisted learning environments.*

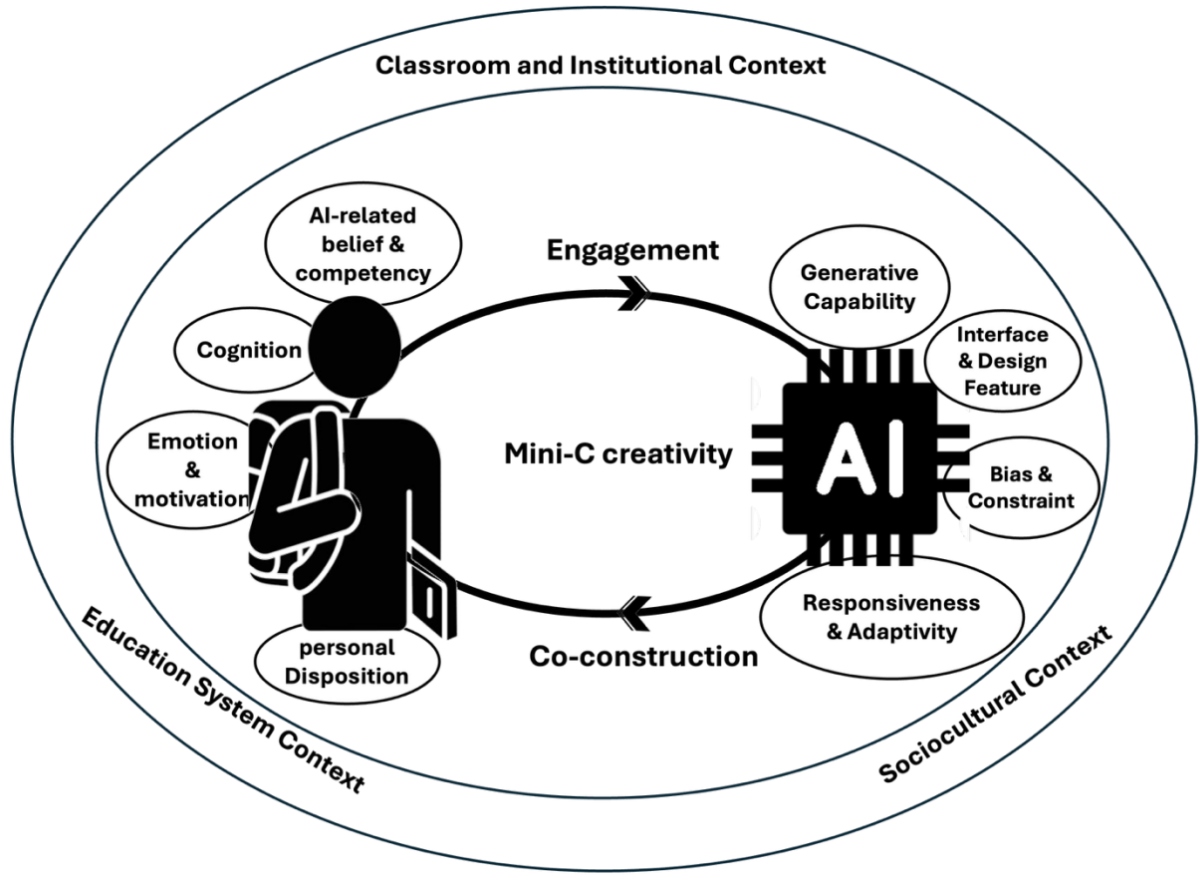

Based on this conceptual model, we propose two focused directions for future research. The first is to examine the process by which students construct mini-c creativity in situated, AI-assisted learning contexts. Creativity, particularly at the mini-c level, is not an isolated event or trait but an evolving process of sense-making, interpretation, and transformation. In AI-assisted learning environments, this creative process is distributed across human and technological agents, shaped by the student's intentional actions and the tool's generative affordances. In this regard, students' agentic engagement—such as initiating prompts, revising inputs, interpreting outputs, and reflecting on outcomes—serves as a key mechanism through which creative meaning is constructed. These actions unfold within specific learning tasks and disciplinary contexts, where the alignment between students' goals, the AI's outputs, and the broader learning environment affects how mini-c creativity takes form.

Studying this process requires fine-grained, temporally sensitive methods that can capture the dynamics of human–AI interaction. Methodologically, researchers might draw on multimodal learning analytics, screen recordings, keystroke logs, and retrospective interviews to trace how creativity is enacted across time. For example, analyzing how students iteratively adjust prompts or evaluate AI-generated alternatives can reveal underlying cognitive and metacognitive strategies tied to creative thinking. Think-aloud protocols and interaction analysis can further expose how students navigate uncertainty, negotiate meaning, and make intentional choices within the AI-mediated space. The significance of this line of research lies in its potential to move beyond surface-level measures of creativity to illuminate how students develop personally meaningful knowledge and insight through interaction with AI. This understanding can guide the design of learning environments that cultivate deep engagement, exploratory thinking, and reflective practices essential to student-driven creativity.

The second research direction focuses on evaluating the outcomes of agentic engagement, specifically in terms of creative performance and the quality of students' outputs in AI-assisted learning. While the process of creativity is vital, it is equally important to assess what results from students' agentic interactions with GenAI tools in authentic, task-rich learning environments. This includes examining whether their engagement leads to work that is original, contextually appropriate, and coherent. Agentic engagement might serve as a predictor of creative success, provided it is strategically and meaningfully enacted. Linking agentic behaviors with observable outcomes enables us to distinguish between surface-level use of AI and more substantive, learner-directed engagement that fosters meaningful, creative development. Additionally, by tracing backward from the quality of creative outputs, researchers can identify which forms and patterns of agentic engagement are most strongly associated with productive and meaningful creative work. This line of inquiry can generate actionable insights for designing pedagogies and intelligent systems that enhance students' creative engagement and learning outcomes.

**A Conclusion and An Opening**

This chapter has explored the evolving landscape of human creativity in AI-assisted learning environments through the conceptual lens of student agency. Drawing on four theoretical perspectives of agency, we have synthesized emerging discourses on how the integration of GenAI influences, enables, and challenges students' creative development. Central to our analysis is the perspective of effortful agency, as creativity and motivation are deeply intertwined, as shown in previous studies. Rather than viewing creativity as a static trait or isolated output, we conceptualize it as a process of meaning-making shaped by students' proactive, intentional, and reflective engagement with AI tools. To extend this line of inquiry, we proposed a new theoretical framework of *agentic engagement in AI-assisted learning environments*, developed through a grounded theory approach. We link this framework to the concept of Mini-c creativity, leading to two key directions for future research on creative process and performance. By doing so, we highlight how creativity in AI-assisted contexts is not simply about generating impressive outputs, but about exercising agency in ways that lead to authentic, self-directed intellectual growth.

While the chapter provides an overview of major arguments in educational discourses, these ideas may be interpreted differently across diverse cultural and geographic regions. Particularly, concepts such as *mini-C creativity*—defined as personally meaningful creative expression—and *student agency*—enacted between individuals and their environments—are deeply shaped by local cultural norms, values, and educational traditions. In some contexts, creativity may be closely linked to collective goals, conformity, or social harmony, rather than individual novelty or self-expression (Goncalo & Staw, 2006). Likewise, agency may be understood not as autonomy or personal choice, but as relational and contextually negotiated within hierarchical or communal structures (Hernandez & Iyengar, 2001). For example, in collectivist cultures, students' agentic actions might center on contributing to group success or aligning with teacher expectations, rather than asserting personal authorship. These cultural orientations influence how students engage with GenAI tools—whether they see such technologies as spaces for personal exploration, shared problem-solving, or guided learning. As such, research on student agency and creativity in AI-assisted environments must attend to how these constructs are locally situated, recognizing that what counts as agentic or creative action may vary significantly across educational and cultural settings.

The theoretical insights presented in this chapter can be translated into pedagogical strategies for educational practitioners. Each of the four traditions of student agency implies distinct approaches to fostering creativity, enabling educators to tailor instructional methods to specific learning goals. For example, the perspective of dynamically emergent agency, which emphasizes coagency, highlights the value of designing activities that support both interpersonal and human–AI co-creativity. Pedagogical activities such as peer-AI co-editing workshops or collective interpretation of GenAI outputs can help students engage in collective sense-making and iterative creation. In addition, integrating the concepts of agentic engagement and mini-c creativity supports the design of a *critique–create–reflect* GenAI prompting strategy. This approach invites students to evaluate or challenge GenAI-generated outputs and reflect on how the outputs relate to their personal meanings and learning goals. Such a strategy encourages critical engagement, personal authorship, and iterative learning, positioning students as active agents in shaping and interpreting creative work in partnership with AI.

    As AI technologies continue to evolve at a rapid pace, so too does the capability of GenAI tools to participate in increasingly complex and creative tasks. What GenAI can generate today—whether in text, image, or multimodal formats—already exceeds the expectations of even a few years ago, and these advancements are accelerating. As these tools become more context-aware, adaptive, and interactive, they are likely to take on even more prominent roles in shaping how students think, create, and learn. In this context of continual change, this chapter is not intended as a definitive conclusion, but rather as an opening—a conceptual starting point and an invitation for ongoing dialogue, inquiry, and collaboration. There are many more questions under-examined: How will future iterations of GenAI redefine the boundaries between human and machine creativity? What new forms of agency will students need to develop to meaningfully engage with these tools? How do we support students in becoming active, reflective, and ethical participants in human–AI creative partnerships? These questions are not fixed, and neither are the answers. They must be revisited as GenAI evolves—and as we collectively imagine the future of creativity in education.